\documentclass[showpacs,preprintnumbers,eqsecnum,twocolumn,prd]{revtex4-1}  

\usepackage{graphicx}
\usepackage{latexsym}
\usepackage{amsmath}
\usepackage{amssymb}

\def\beq{\begin{equation}}
\def\eeq{\end{equation}}
\begin{document}

\title{Detweiler's gauge-invariant redshift variable: analytic determination of the nine and nine-and-a-half post-Newtonian self-force contributions}

\author{Donato \surname{Bini}$^1$}
\author{Thibault \surname{Damour}$^2$}

\affiliation{$^1$Istituto per le Applicazioni del Calcolo ``M. Picone'', CNR, I-00185 Rome, Italy\\
$^2$Institut des Hautes Etudes Scientifiques, 91440 Bures-sur-Yvette, France}

\date{\today}
\begin{abstract}
Continuing our analytic computation of  the first-order self-force contribution to  Detweiler's redshift variable we provide the exact
expressions of the ninth and ninth-and-a-half post-Newtonian terms.
\end{abstract}
\maketitle

\section{Introduction}

The prospect of soon detecting the gravitational-wave signals emitted by coalescing compact binaries motivates a renewed study of the  general relativistic two-body problem.
One of the useful lines of attack on this problem is  the gravitational self-force program, which considers large-mass-ratio binary systems ($m_1\ll m_2$), and uses an
expansion in powers of the mass ratio $q\equiv m_1/m_2 \ll 1$.  Within this program, Detweiler \cite{Detweiler:2008ft} has emphasized the importance of focussing
on the computation of gauge-invariant quantities, and he gave (for the case of circular motions) the example of the function relating the redshift $U^t=dt/ds$ along the worldline of 
the small mass $m_1$ to the orbital frequency $\Omega$.  To first order in $q$ this gives rise to the gauge-invariant function $U_1^t(y)$, where 
$U^t(y)= (1-3y)^{-1/2} + q \, U_1^t(y) +  O(q^2)$ and where $ y \equiv (G m_2 \Omega/c^3)^{2/3} $ denotes a dimensionless parameter related
to the orbital frequency. 
Note that $y$ can be considered as measuring (in a gauge-invariant way) the dimensionless gravitational potential $ G m_2 /(c^2 R_{\Omega}) $, with $R_{\Omega}$
denoting the invariant radius canonically associated with  $\Omega$ via Kepler's law around the large mass: $G m_2 = \Omega^2   R_{\Omega}^3$.  In the following, we shall denote the
mathematical argument of the first-order self-force function $U_1^t(.)$ by $u$ (evoking a gravitational potential) rather than $y$. [This notation is purely a matter of choice.]
We henceforth also often set $G=c=1$.

Detweiler \cite{Detweiler:2008ft} has shown that $U_1^t(u)$ could be computed in terms of the first-order metric perturbation of  a Schwarzschild metric of mass $m_2$,
say   $\delta g_{\mu\nu}= g_{\mu\nu} (x^\lambda ;m_1,m_2) - g_{\mu\nu}^{\rm Schw} (x^\lambda ;m_2) \equiv q \, h_{\mu\nu}(x^\lambda) + O(q^2)$ as
\beq
\label{Detw}
U^t_{1}(u)=\frac1{2(1-3u)^{3/2}}h_{kk}^R(u) \,,
\eeq
where
\beq
\label{h_kk_reg}
 h_{kk}^{R}(u) := \left[ h_{\mu\nu}(x^\lambda)\right]^{R} k^{\mu} k^{\nu} \,.
 \eeq
Here $k^{\mu}$ denotes  the Killing vector $k= \partial_t + \Omega \partial_{\varphi}$,   and the superscript  $R$
 denotes the   {\it regularized}  value of  $h_{\mu\nu}(x^\lambda)$ on the world line of the small mass $m_1$.  [When evaluating  the first-order
 quantity $h_{kk}^{R}$ along a circular orbit of (coordinate) radius $R_0$ it is enough to use the approximation $ R_0 \approx R_{\Omega} = m_2/u$.]
 
 The beginning of the post-Newtonian (PN) expansion
of the (first-order) self-force contribution $U_1^t(u)$ was analytically derived in Ref. \cite{Detweiler:2008ft}, namely :
\beq
U_1^t(u)= - u - 2 u^2 -5 u^3 - \cdots
\eeq
Here,  the first term ($-u$) is of Newtonian order, so that the second ($- 2 u^2 $) and third ($ -5 u^3$) respectively represent  1PN and 2PN contributions.
More generally, a term $\propto u^{n+1}$ corresponds to the $n$PN level in $U_1^t(u)$. 

The 3PN term was analytically derived (using full PN theory)  by Blanchet et al. \cite{Blanchet:2009sd}. In 2013, 
 we \cite{Bini:2013zaa} showed how to analytically compute the 4PN term by a combined use of Regge-Wheeler-Zerilli (RWZ) formalism for the Schwarzschild perturbations together with the hypergeometric-expansion analytical solutions of the RWZ radial equation obtained by  Mano, Suzuki and Takasugi \cite{Mano:1996mf} (MST). We then progressively extended the analytical knowledge of the PN expansion of $h_{kk}^R$ up to the 8.5 PN level \cite{Bini:2013rfa, Bini:2014nfa}. [See the latter works
 for references to other related analytical studies.]
 
 Parallely to these analytical studies, Detweiler's redshift variable was numerically computed in Refs.  \cite{Detweiler:2008ft,Blanchet:2009sd,Akcay:2012ea},
 and these numerical data were used to extract numerical estimates of several  higher-order (then unknown) PN expansion coefficients  \cite{Detweiler:2008ft,Blanchet:2009sd}.
  A breakthrough in this
 extraction of PN coefficients from numerical self-force calculations was accomplished by  Shah,  Friedman and Whiting \cite{Shah:2013uya} who numerically
 evaluated the MST hypergeometric-expansion of  $U_1^t(u)$  to one part in $10^{225}$ for orbital radii extending up to $10^{30} Gm_2/c^2$ . This extremely high numerical accuracy  on  $U_1^t(u)$ for extremely small values of the argument $u$ allowed them to numerically
 extract PN coefficients up to the 10.5 PN level, and also to provide educated guesses for the exact
 analytical form of several high-order PN coefficients. 
 
 We have shown in \cite{Bini:2013rfa, Bini:2014nfa}  that the results of  Shah,  Friedman and Whiting \cite{Shah:2013uya} agreed with our (fully) analytical
 ones up to the highest PN level we had then computed, namely the 8.5PN level. The aim of the present short note is to report on an extension of our analytical
computation to the 9.5 PN level (using the techniques explained in our previous papers), and on its comparison with the results of Shah et al.
  
\section{New terms in the PN expansion of  $U_1^t(u)$ at the 9 and 9.5 PN levels}

Following the notation   of Eq. (21) in  Ref. \cite{Shah:2013uya}, we write
\begin{eqnarray}
U_1^t(u)&=&U_1^t(u)|_{8.5\rm PN}+(\alpha_9 -\beta_9 \ln u +\gamma_9 \ln^2 u)u^{10}\nonumber\\
&& +(\alpha_{9.5}-\beta_{9.5}\ln u)u^{21/2}\,,
\end{eqnarray}
where $U_1^t(u)|_{8.5\rm PN}$ is known from \cite{Bini:2014nfa}.  Here ($\alpha_9$,  $\beta_9$, $\gamma_9$), and  ($\alpha_{9.5}$, $\beta_{9.5}$) are the coefficients of,
respectively, the 9PN ($u^{10}$) and 9.5PN  ($u^{21/2}$) terms, which we have now analytically derived. Our results for these coefficients are:
\begin{widetext} 
\begin{eqnarray}
\alpha_9&=& -\frac{10480362137370508214933}{2044301131372500}+\frac{5921855038061194}{442489422375}\gamma-\frac{2076498568312502}{442489422375}\ln(2)\nonumber\\
&&+\frac{10221088}{2835}\zeta(3)-\frac{16110330832}{9823275}\gamma^2+\frac{32962327798317273}{549755813888}\pi^4\nonumber\\
&& -\frac{11665762236240841}{226072985600}\pi^2+\frac{33662546992}{9823275}\ln(2)^2-\frac{27101981341}{100663296}\pi^6\nonumber\\
&& -\frac{58533203125}{15567552}\ln(5)-\frac{96889010407}{277992000}\ln(7)+\frac{9647042994387}{392392000}\ln(3)-\frac{94770}{49}\ln(3)^2\nonumber\\
&& -\frac{10020829088}{9823275}\ln(2)\gamma-\frac{189540}{49}\ln(2)\ln(3)-\frac{189540}{49}\gamma\ln(3)\nonumber\\
&=&\boxed{ -32239.62759509255641236770603459209615}2340561204299\nonumber\\
\beta_9 &=&\frac{16110330832}{9823275}\gamma-\frac{2921280466785797}{442489422375}+\frac{5010414544}{9823275}\ln(2)+\frac{94770}{49}\ln(3)\nonumber\\
&=& \boxed{-3176.9291811539692063923388326926660888}222379193868\nonumber\\
\gamma_9 &=& -\frac{4027582708}{9823275}\nonumber\\
\alpha_{9.5}&=&-\frac{30185191523470507}{12236744520000}\pi-\frac{410021764}{385875}\ln(2)\pi-\frac{198373004}{1157625}\pi\gamma-\frac{1055996}{11025}\pi^3+\frac{246402}{343}\pi\ln(3)  \nonumber\\
&=& \boxed{-10864.6255867062440752457674}43506686658105844986920\nonumber\\
\beta_{9.5}&=& \frac{99186502}{1157625}\pi 
\end{eqnarray}
\end{widetext}
Our analytically derived results for $\gamma_9$ and $\beta_{9.5}$  agree with the numerical-based analytical expressions previously obtained for
these two particular coefficients by  Shah et al. \cite{Shah:2013uya}. Concerning the other (newly analytically computed)
coefficients, namely   $\alpha_9$,  $\beta_9$, and $\alpha_{9.5}$  we have indicated by boxes in the above equations the extent to which our results
agree with the numerical estimates given by  Shah  et al. \cite{Shah:2013uya}. More precisely, the boxes above include one more digit than those given in
Table I of  \cite{Shah:2013uya}. In all cases, the agreement is perfect modulo possible rounding effects on the last digit quoted in  \cite{Shah:2013uya}.

\section{Concluding remarks}
The  analytic computation of the post-Newtonian expansion of  the first-order self-force contribution $U_1^t(u)$ to  Detweiler's redshift function $U^t(\Omega)$
 has been raised here to the nine and nine-and-a-half PN level, thereby providing the exact analytical expressions of terms which were previously obtained only 
 numerically by Shah et al. \cite{Shah:2013uya}. 
 
 Let us finally note that, using the results of Refs. \cite{LeTiec:2011ab,Barausse:2011dq}, our results can
 be translated into the computation of the nine and nine-and-a-half PN contributions to the linear-in-mass-ratio piece of
 the main radial potential $A(u; \nu)$ of the effective one-body formalism  \cite{Buonanno:1998gg,Damour:2000we}. Denoting,
  $A(u; \nu) = 1- 2 u + \nu a(u) + O(\nu^2)$  (where $\nu \equiv   m_1 m_2 /(m_1+m_2)^2 = q/(1+q)^2$), we have
\begin{eqnarray}
a(u)&=& a_{8.5\rm PN}(u)+(a_{10}^c +  a_{10}^{\ln} \ln u + a_{10}^{\ln^2} \ln^2 u) u^{10}\nonumber\\
&& + (a_{10.5}^c + a_{10.5}^{\ln} \ln u) u^{21/2}\,,
\end{eqnarray}
where $ a_{8.5\rm PN}(u)$ was given in  \cite{Bini:2014nfa}, and where the newly derived 9PN and 9.5PN coefficients are:
\begin{widetext}
\begin{eqnarray}
a_{10}^c &=& \frac{18605478842060273}{7079830758000}\ln(2)-\frac{1619008}{405}\zeta(3)-\frac{21339873214728097}{1011404394000}\gamma\nonumber\\
&&+\frac{27101981341}{100663296}\pi^6-\frac{6236861670873}{125565440}\ln(3)+\frac{360126}{49}\ln(2)\ln(3)+\frac{180063}{49}\ln(3)^2\nonumber\\
&& -\frac{121494974752}{9823275}\ln(2)^2-\frac{24229836023352153}{549755813888}\pi^4+\frac{1115369140625}{124540416}\ln(5)+\frac{96889010407}{277992000}\ln(7)\nonumber\\&& +\frac{75437014370623318623299}{18690753201120000}-\frac{60648244288}{9823275}\ln(2)\gamma+\frac{200706848}{280665}\gamma^2\nonumber\\
&& +\frac{11980569677139}{2306867200}\pi^2+\frac{360126}{49}\gamma\ln(3)\nonumber\\
a_{10}^{\ln}&=& -\frac{21275143333512097}{2022808788000}+\frac{200706848}{280665}\gamma-\frac{30324122144}{9823275}\ln(2)+\frac{180063}{49}\ln(3)\nonumber\\
a_{10}^{\ln^2}&=& \frac{50176712}{280665}\nonumber\\
a_{10.5}^c&=& -\frac{185665618769828101}{24473489040000}\pi+\frac{377443508}{77175}\ln(2)\pi+\frac{2414166668}{1157625}\pi\gamma -\frac{5846788}{11025}\pi^3-\frac{246402}{343}\pi\ln(3)\nonumber\\
a_{10.5}^{\ln} &=& \frac{1207083334}{1157625}\pi \,.
\end{eqnarray}
The corresponding numerical values are (consistently with, but more accurately than in,
 Eqs. (27), (28) in \cite{Bini:2014nfa})
\begin{eqnarray}
a_{10}^c &=&4845.87055 70194 44177 34739 34215 79822 17665 62229 29365\ldots  \nonumber\\ 
a_{10}^{\ln}&=& 
-8207.44191 51719 61061 49591 36719 85378 06462 24938 90134\ldots \nonumber\\ 
a_{10}^{\ln^2}&=& 178.77794 52372 04496 46375 57230 14982 27424 15335 00792\ldots \nonumber\\ 
a_{10.5}^c&=& -28324.30746 52136 28065 67119 45153 96169 33632 87226 51715\ldots \nonumber\\
a_{10.5}^{\ln} &=& 3275.81395 90671 19914 18131 48554 51516 85158 78070 63565\ldots\,. \end{eqnarray}
\end{widetext}

\subsection*{Acknowledgments}
  
D.B. thanks the Italian INFN (Naples) for partial support and IHES for hospitality during the development of this project.
Both  authors are grateful to ICRANet for partial support.

\end{document}